\documentclass[12pt,a4paper,tikz]{article}

\usepackage{bm}
\usepackage{bbold}
\usepackage{amsmath}
\usepackage{eqnarray}
\usepackage{amssymb}
\usepackage{natbib}
\usepackage{makeidx}
\usepackage[usenames, dvipsnames]{color}
\usepackage[bbgreekl]{mathbbol}
\DeclareSymbolFontAlphabet{\amsmathbb}{AMSb}%
\DeclareSymbolFontAlphabet{\mathbb}{AMSb}
\usepackage{amsfonts}
\usepackage{mathbbol}
\usepackage{txfonts}

\usepackage{graphics}
\DeclareGraphicsExtensions{.png,.pdf}
\usepackage{epsfig}
\usepackage{epstopdf}
\usepackage{ragged2e}
\usepackage{hyperref}

\usepackage{ccite}

\usepackage[utf8]{inputenc}
\usepackage[english]{babel}

\usepackage{geometry}
\usepackage{fancyhdr}
\usepackage{setspace}
\usepackage{multicol}

\begin{document}

\raggedright
\huge
Astro2020 Science White Paper \linebreak

Understanding the evolution of close white dwarf binaries \linebreak
\normalsize

\noindent \textbf{Thematic Areas:} \hspace*{60pt} $\square$  Planetary Systems \hspace*{10pt} $\square$ Star and Planet Formation \hspace*{20pt}\linebreak
$\square$ \hspace*{-9.5pt}x Formation and Evolution of Compact Objects \hspace*{31pt} $\square$ Cosmology and Fundamental Physics \linebreak
  $\square$ \hspace*{-9.5pt}x Stars and Stellar Evolution \hspace*{1pt} $\square$ Resolved Stellar Populations and their Environments \hspace*{40pt} \linebreak
  $\square$    Galaxy Evolution   \hspace*{45pt} $\square$             Multi-Messenger Astronomy and Astrophysics \hspace*{65pt} \linebreak
  
\textbf{Principal Author:}

Name:	Odette Toloza
 \linebreak						
Institution:  University of Warwick
 \linebreak
Email: odette.toloza@warwick.ac.uk

\justify
\textbf{Co-authors:} Elm{\'e} Breedt	(University of Cambridge),
Domitilla De Martino (INAF-OA-Naples),
Jeremy Drake (CfA),
Alessandro Ederoclite (Universidade de S\~ao Paulo),
Boris G{\"a}nsicke (University of Warwick),
Matthew Green (University of Warwick),
Jennifer Johnson (Ohio State University)
Christian Knigge (University of Shouthampton),
Juna Kollmeier (Carnegie Observatories)
Thomas Kupfer (KITP),
Knox Long (STScI),
Thomas Marsh (University of Warwick),
Anna Francesca Pala (ESO),
Steven Parsons (University of Sheffield),
Tom	Prince (CALTECH),
Roberto Raddi (Universt\"at Erlangen-N\"urnberg),
Alberto Rebassa-Mansergas (Universitat Polit{\`e}cnica de Catalunya),
Pablo Rodr{\'i}guez-Gil (Instituto de Astrof{\'i}sica de Canarias/Universidad de La Laguna),
Simone Scaringi (Texas Tech University),
Linda Schmidtobreick (ESO),
Matthias Schreiber (Universidad de Valpara{\'i}so),
Ken Shen (UC Berkeley),
Danny Steeghs (University of Warwick),
Paula Szkody (University of Washington),
Claus Tappert (Universidad de Valpara{\'i}so),
Silvia Toonen (University of Birmingham),
Axel Schwope (AIP),
Dean Townsley (University of Alabama),
Monica Zorotovic (Universidad de Valpara{\'i}so)

\vspace{0.5cm}
\textbf{Abstract  (optional):} Interacting binaries containing white dwarfs can lead to a variety of outcomes that range from powerful thermonuclear explosions, which are important in the chemical evolution of galaxies and as cosmological distance estimators, to strong sources of low frequency gravitational wave radiation, which makes them ideal calibrators for the gravitational low-frequency wave detector \textit{LISA} mission. However, current theoretical evolution models still fail to explain the observed properties of the known populations of white dwarfs in both interacting and detached binaries. Major limitations are that the existing population models have generally been developed to explain the properties of sub-samples of these systems, occupying small volumes of the vast parameter space, and that the observed samples are severely biased. The overarching goal for the next decade is to assemble a large and homogeneous sample of white dwarf binaries that spans the entire range of evolutionary states, to obtain precise measurements of their physical properties, and to further develop the theory to satisfactorily reproduce the properties of the entire population. While ongoing and future all-sky high- and low-resolution optical spectroscopic surveys allow us to enlarge the sample of these systems, high-resolution ultraviolet spectroscopy is absolutely essential for the characterization of the white dwarfs in these binaries. The \textit{Hubble Space Telescope} is currently the only facility that provides ultraviolet spectroscopy, and with its foreseeable demise, planning the next ultraviolet mission is of utmost urgency.

\pagebreak

\justify

\textbf{\large Astrophysical context}
\medskip\\
\noindent White dwarfs in close binaries include a vast range of exotic systems such as white dwarf pulsars \citep{Marsh2016AStar} or the ultra-compact AM\,CVns. They can lead to powerful thermonuclear explosions, and are strong sources of low-frequency gravitational waves that will serve as calibrators for the Laser Interferometer Space Antenna. While binaries containing white dwarfs represent the simplest case of the evolution of compact binaries, there remain discrepancies between the current population models and the properties of the observed samples \citep[e.g.][]{Zorotovic2011Post-common-envelopePre-CVs, Pala2017EffectiveEvolution}. Among the most important aspects of compact binary evolution are the multiple pathways leading to a Type Ia Supernova (SN\,Ia). The discovery of a range of related thermonuclear explosions, such as SN\,Iax \citep{Foley2012TypeExplosion} and the calcium-rich transients \citep{Perets2010The2005cz}, as well as over- \citep{Filippenko1992TheDwarf} and under-luminous \citep{Filippenko1992The4374} SN\,Ia challenges our understanding of the evolution of close binaries containing white dwarfs. 
\noindent Furthermore, accreting white dwarf binaries are versatile laboratories for accretion physics in extreme conditions, serving as test benches for models of more complex astrophysical systems, including active galactic nuclei or quasars \citep{Pringle1981AccretionAstrophysics}.

\begin{figure}[h]
   \centering
       \parbox{0.37\textwidth}{\includegraphics[width=0.40\columnwidth]{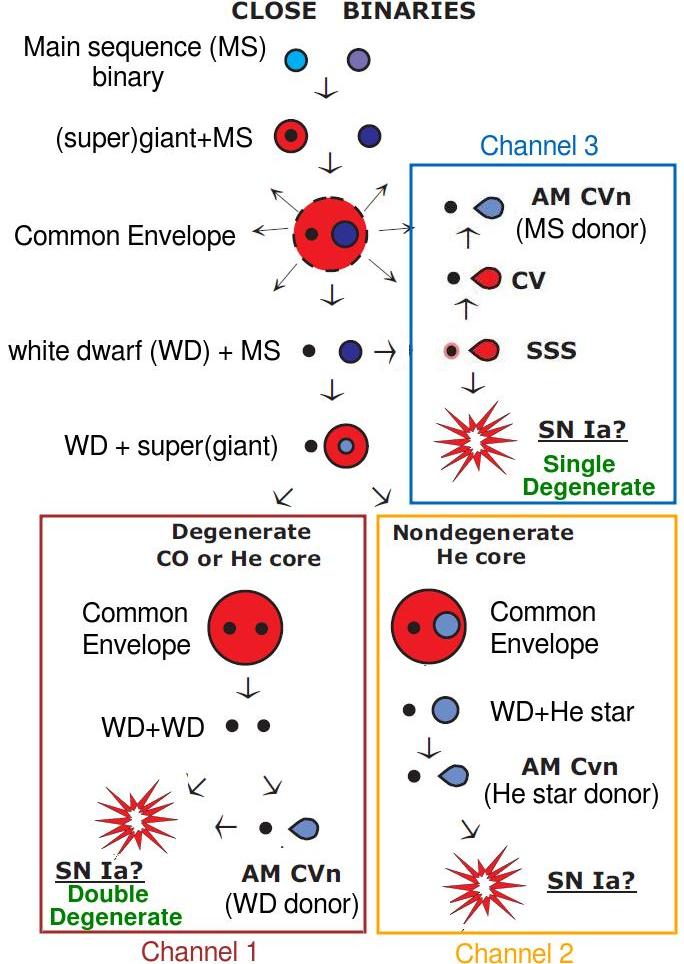}\textit{\caption{\small Evolution of close binary, resulting in mergers or systems with high- (SSSs) and low accretion rates (CVs and AM\,CVns).} \label{fig:schema}}}
    \begin{minipage}{0.62\textwidth}
Fig. \ref{fig:schema} illustrates various evolutionary outcomes of a pair of main sequence stars. The physics at the different stages along the evolution is complex, and open questions are:

\begin{itemize}
\setlength\itemsep{-0pt}
    \item What are the sources of energy that expel the common envelope, and what fraction of the available energy is used?
    \item What are the physical properties of the outcomes of the  first (and second) common envelope?
    \item Can angular momentum loss, which drives the evolution, be described with a single prescription of magnetic braking?
    \item What physics is important in accretion discs, winds, and jets?
    \item Do white dwarf binaries evolve through (a) phase(s) of stable or unstable mass transfer?
    \item What is the fraction of systems that evolve through the different channels?
    \item Do the majority of donors in SN\,Ia progenitors have a degenerate nature? Or non-degenerate?\\
\end{itemize}

    \end{minipage}
\end{figure}

\noindent Progress in answering these questions has been limited since (1) the currently known samples of white dwarf binaries are incomplete and severely affected by selection biases, and (2) only a very small fraction of the $\simeq$1500  white dwarf binaries known has been accurately characterized. In particular, the physical properties of the white dwarfs are crucial for the further development of the theory of close binary evolution. Detailed studies of the physical and atmospheric properties of (non-)interacting white dwarfs in close binaries are \textit{only} possible in the ultraviolet since the contamination from the companion and/or accretion disc are minimal at these short wavelengths. \textit{Therefore, high-resolution ultraviolet spectroscopy is critically important for our understanding of compact binary evolution, with ramifications for a wide range of astrophysics}.

\clearpage

\noindent
\textbf{\large Key aspects of white dwarf binary  evolution}
\smallskip\noindent
    
\begin{figure}
\centering
\includegraphics[width=1.0\columnwidth]{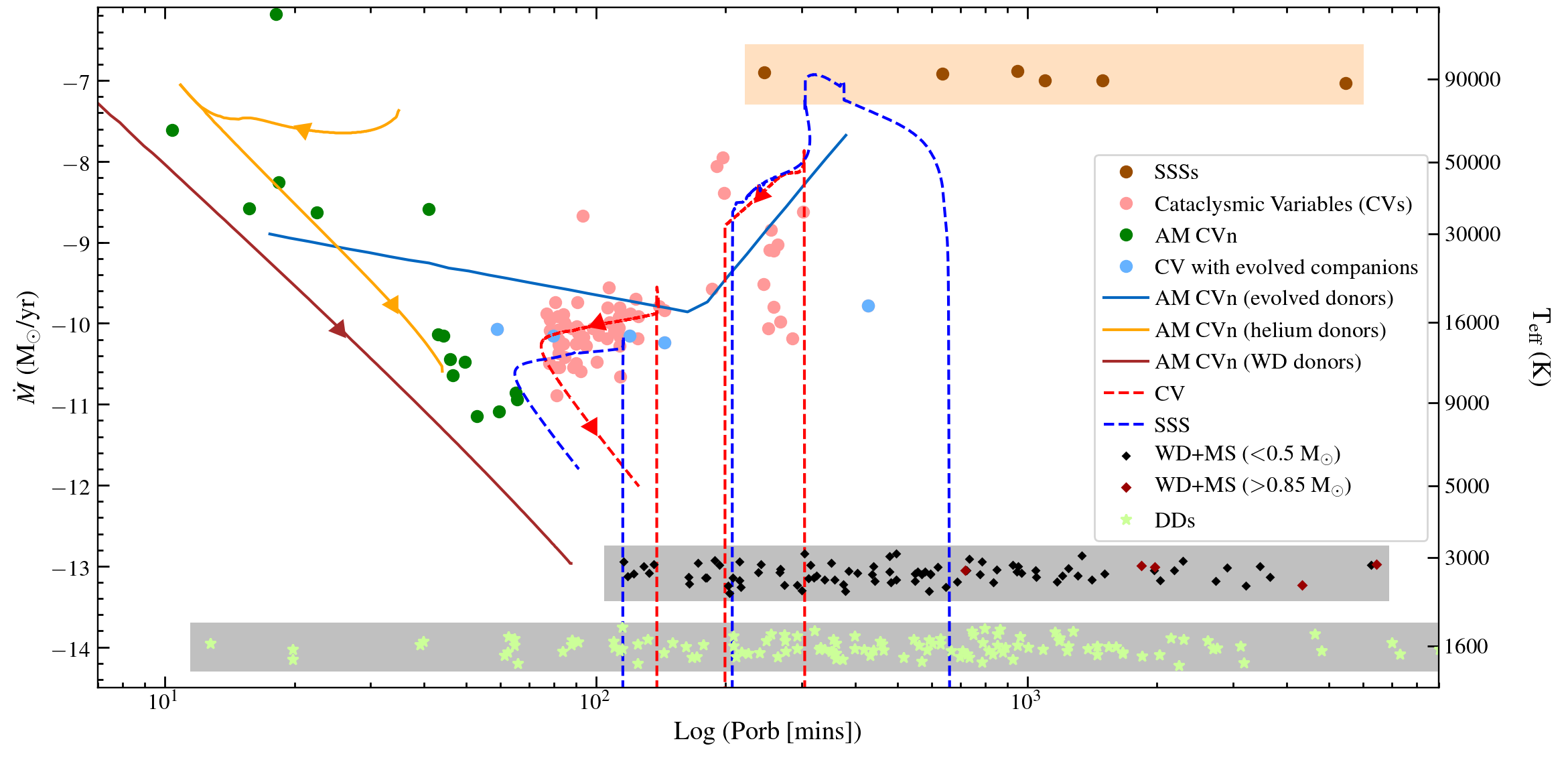}
\textit{\caption{\small Theoretical evolutionary tracks (lines) and observed populations of accreting white dwarfs (dots) and detached systems (diamonds and stars). \textbf{\textit{Accreting white dwarfs:}} the accretion rate (left Y-axis) is a key prediction of the evolution models (solid and dashed tracks), and together with the white dwarf mass determines its effective temperature, which is measured from ultraviolet spectroscopy (right Y-axis, for a fixed white dwarf mass of M$_{\mathrm{WD}}$=0.83\,M$_{\odot}$). The accretion rates for cataclysmic variables (CVs) with low mass companions (pink dots, \citealt{Townsley2009CataclysmicLoss, Pala2017EffectiveEvolution}) and those with nuclearly evolved companions (blue dots, \citealt{Pala2017EffectiveEvolution}) have been determined from fits to  \textit{HST}/STIS and \textit{HST}/COS spectra. For guidance, we show the crude estimates of the accretion rates determined for AM\,CVns using \textit{Gaia} DR2 parallaxes combined with multi-band photometric data (green dots, \citealt{Ramsay2018TheDR2})~--~and stress that the \textit{GALEX} photometry is not sufficient to model the white dwarf temperature. The three evolutionary channels of AM\,CVns are over-plotted with solid lines, with the donors being white dwarfs (brown), helium-stars (orange), or evolved main sequence stars (blue). Evolutionary tracks are aso shown for systems containing a white dwarf with M$_{\mathrm{WD}}$=0.8\,M$_{\odot}$ which evolve either as canonical cataclysmic variables (red) or undergo hydrogen shell burning (blue). Both tracks overlap with the observed CVs and those systems that contain evolved companions, and hence only ultraviolet N/C flux ratios can disentangle the two populations. The top region shows the super-soft X-ray sources (SSSs) for which their orbital periods are known (brown dots, \citealt{Kahabka1997LUMINOUSSOURCES}). They are enclosed by a pale orange box defining the accretion rates needed to sustain hydrogen burning on the white dwarf surface. \textbf{\textit{Detached systems:}} The grey boxes near the bottom show the period distributions of detached post common envelope binaries containing a low mass M-type companion ($<$0.5\,M$_{\odot}$, black diamonds) and more massive ones ($>$0.85\,M$_{\odot}$, dark red diamonds), and the double white dwarfs (bright green stars). The systems are offset randomly by small amounts in accretion rate for clarity~--~in reality, post-common envelope binaries can accrete from the winds of their companions at $\simeq10^{-16}$ to $10^{-15}\,\mathrm{M_\odot/yr}$}.  \label{fig:figure}}
\end{figure}

\textbf{The common envelope:} Detailed hydro-dynamical simulations of this brief but extremely important phase are still limited in scope \citep{Ivanova2012CommonForward}, and therefore simple equations of energy conservation, parameterizing the efficiency of the common envelope as $\alpha$, are used in binary population models. The value of $\alpha$ has a large impact on the predicted numbers and properties of compact binary populations. The characteristics of observed post common envelope white dwarf plus M-star binaries (black diamonds in Fig. \ref{fig:figure}) are successfully reproduced with $\alpha\simeq0.2-0.3$ \citep{Zorotovic2010Post-common-envelopeEfficiency, Toonen2013TheBinaries, Camacho2014MonteSample}, but larger values are needed to explain post common envelope binaries containing helium white dwarfs \citep{Nelemans2000ReconstructingSpiral-in}. Moreover, additional sources of energy, such as recombination in the envelope, are required to explain the evolution of systems at long orbital periods and with massive companions  \citep[e.g. IK\,Peg \& KOI-3278, ][]{Zorotovic2014MonteEnergy}. An accurate knowledge of the white dwarf mass is important to test the analytical prescriptions of the common envelope. Systems emerging from the common envelope can experience a second common-envelope phase and become ultra-short orbital period binaries (i.e. AMCVn stars and double degenerates). In another evolutionary pathway, the separation of the emerging systems decreases further via angular momentum loss until the secondary overflows its Roche lobe. Through stable mass transfer, matter is accreted onto the white dwarf, forming a cataclysmic variable (CV)


\textbf{Cataclysmic variables}: Angular momentum loss and the resulting mass-transfer rate govern the evolution of CVs (pink dots in Fig. \ref{fig:figure}). The predictions of traditional population models assuming that magnetic braking \citep{Verbunt1969MagneticBinaries} dominates at long orbital periods, and gravitational wave radiation at short periods \citep{Paczynski1967GravitationalBinaries} conflict in several aspects with the observations. \citet{Schreiber2015ThreeEvolution} showed that accounting for the frictional angular momentum loss due to shell flashes on the white dwarf can solve some of these discrepancies: (1) the orbital period distribution \citep{Knigge2006TheVariables}, (2) the space density of CVs \citep{Patterson1998LateVariables, Pretorius2011TheVariables}, and (3) the average white dwarf mass \citep{Zorotovic2011Post-common-envelopePre-CVs}. However, even the updated theory fails to correctly predict the mass transfer rates of CVs at longer periods ($>$3\,h), and for those with nuclearly evolved companions \citep{Pala2017EffectiveEvolution}. Accretion results in compressional heating of the white dwarf \citep{Townsley2009CataclysmicLoss}, hence measuring the white dwarf temperatures and masses in systems spanning a wide range in evolutionary parameter space can constrain and calibrate the angular momentum loss prescriptions.

\textbf{Supersoft X-ray sources}:  In CVs the accretion rate is low ($\lesssim$10$^{-7}$\,M$_{\odot}$/yr) and the white dwarf undergoes intermittent shell flashes. However, at higher rates, $\gtrsim$10$^{-7}$\,M$_{\odot}$/yr, steady hydrogen shell burning is sustained on the white dwarf surface, making it an intense source of soft X-rays (brown dots in Fig. \ref{fig:figure}). These systems are potential SN\,Ia or SN\,Iax progenitors through the single degenerate channel, but have been extremely difficult to characterize observationally because of the self-shielding effect of the X-ray emission \citep{Nielsen2012ObscurationProgenitors}. An alternative approach is to study the systems that evolve through this channel, but failed to reach the ignition mass for a thermonuclear explosion. The high accretion rates in supersoft X-ray sources require donor stars with initial masses $\gtrsim 1.2\,M_\odot$, and therefore progenitors of supersoft X-ray sources are detached white dwarf plus F/G main sequence stars (dark red diamonds in Fig. \ref{fig:figure}), which are powered by the CNO cycle. If the mass ratio comes close to unity before the white dwarf reaches the Chandrasekhar limit, the mass transfer rate drops and the systems then resemble stably accreting white dwarfs which were born with lower mass companions, i.e. CVs. The only signature to distinguish a system that underwent the hydrogen-shell burning phase among the more numerous population of CVs is the fact that the accreting material carries the fingerprint of the CNO cycle. Indeed, about a dozen stably accreting white dwarfs present anomalous enhancement of nitrogen and depletion of carbon, identified by their large ultraviolet N\,\textsc{v}~1240\,\AA/~C\,\textsc{iv}~1550\,\AA\ flux ratios \citep[blue dots in Fig. \ref{fig:figure},][]{Schenker2002AEBinaries, Gansicke2003AnomalousHST/STIS, Sanad2011TwoVariables}. 

\textbf{Accretion physics}: There is no standard model for the accretion process yet. Accretion is expected to occur through a boundary layer, where the particles are decelerated from Keplerian velocities in the disk to the white dwarf stellar rotation \citep{Lynden-Bell1974TheVariables}, and will evaporate if the accretion rate is too low. In addition, the P-Cygni profiles seen in the spectra of high mass accretion rate CVs unambiguously establish the presence of disk winds, likely in a biconical outflow \citep{Sion2004CompositeCygni}, though it is unclear if they originate at the boundary layer or in the inner regions of the disk. Once sufficient matter has built up, the disk undergoes an outburst flushing material faster towards the white dwarf. Then, because of the increased accretion rate  ($\sim$10$^{-10}$M$_{\odot}$/yr), the boundary layer becomes optically thick with temperatures reaching 200\,000--500\,000\,K \citep{Pringle1977SoftNovae}. This boundary layer comprises an important luminosity component in accretion disk. The observed strength of the He\,\textsc{ii} recombination lines (e.g. He\,\textsc{ii} at 1640\,\AA) can be used to deduce the number of He+ ionizing photons emitted by the boundary layer/accretion disk \citep{Hoare1991Boundary-layerVariables}, and therefore place constraints on the temperature of the boundary layer. The boundary layer plays a crucial role in determining the ionization state of the wind, therefore modelling the ultraviolet resonance lines (C\,\textsc{iv}\,1550\,\AA, Si\,\textsc{iv}\,1400\,\AA\ and N\,\textsc{v}\,1240\,\AA) provides constraints on the geometry and velocity of the outflow. 

\textbf{AM\,CVn stars}: The white dwarfs in these ultra-compact binaries accrete from a low-mass helium-dominated donor. They are the final outcomes of the evolution involving one or two common envelopes leading to configurations with ultra-short orbital periods down to a few minutes~--~making them powerful low-frequency gravitational wave beacons. Fig. \ref{fig:schema} shows the three formation channels that are thought to contribute to the population of AM\,CVns (green dots in Fig. \ref{fig:figure}): (1) mass transfer between two white dwarfs, where the lower-mass donor star is helium-rich \citep{Paczynski1967GravitationalBinaries}, (2) accretion from a low-mass non-degenerate helium donor \citep{IbenIcko1987EvolutionaryBinaries}, or (3) accretion from the stripped core of a nuclearly evolved donor which possibly underwent a supersoft X-ray phase \citep{Podsiadlowski2003CataclysmicStars}. The relative fractions of the individual channels are disputed \citep{Shen2015EveryMerge, Ramsay2018TheDR2}, and poorly constrained by observations.  A key method to distinguish between the different birth channels of AM\,CVns requires abundance measurements; in particular, the ratios of N/C and N/O can constrain the initial mass of the donor \citep{Nelemans2010TheFormation}.

\textbf{Double degenerates}: The outcome of two common envelopes also leads to detached short-period double degenerates (green stars in Fig. \ref{fig:figure}), some of which are likely SN\,Ia progenitors (Fig.\,\ref{fig:schema}). While  theoretical models have explored in detail the white dwarf ignition \citep{Webbink1984DoubleSupernovae, Livio2003HaveDiscovered, Guillochon2010SurfaceInstabilities}, so far only one pair of white dwarfs that will merge within a Hubble time \textit{and} has a sufficiently large mass that it \textit{may} ignite as SN\,Ia has been found \citep{Maxted2000KPD1930+2752Progenitor}. 
The optical spectra of many known double-degenerates are dominated by just one of the white dwarfs, limiting the amount of system properties that can be measured. However, for pairs with unequal masses and temperatures, the hotter and cooler components dominate at ultraviolet and optical wavelengths, respectively. Hence, the properties of the hotter component can be disentangled via ultraviolet spectroscopy \citep[e.g.][]{Bours2015AOrigin, Fusillo-Gentile2017CanWD2105-820}.

\textbf{Supernova survivors}: The field of thermonuclear supernovae underwent a recent breakthrough with the discovery of hypervelocity stars that constitute the first direct evidence for two distinct classes of supernova survivors ejected from the binaries: the former white dwarf donors of SN\,Ia \citep{Shen2018}  and the partly burned white dwarf accretor of either a SN\,Iax \citep{Vennes2017} or an electron-capture supernovae \citep{Raddi2019ASupernovae}. The accurate measurement of the chemical composition of three of latter exotic systems shows that they are ONe-dominated atmosphere white dwarfs, which are enriched with remarkably similar amounts of nuclear ashes of partial O- and Si-burning \citep{Raddi2019ASupernovae}. Accurate measurements of the abundances, temperatures, and masses of these stars will provide insight into their evolution and their explosion mechanisms, supernova rates, and nucleosynthetic yields.

\medskip
\textit{Our ultimate goal is to identify and accurately measure the parameters of a sufficient number of all the different types of systems outlined above to fully map the entire parameter space shown in Fig. \ref{fig:figure}, which is essential to achieve a proper understanding of the evolution of compact binaries}. Below we outline the required facilities.

\newpage
\noindent
\textbf{\large All-sky optical spectroscopic surveys to identify white dwarf binaries}\\
While short (and high) time-domain cadence imaging surveys are biased towards short term-variability systems, spectroscopy has proven to be unaffected by observational selection effects in the identification of white dwarf binaries, most effectively demonstrated by SDSS (\citealt{Gansicke2004ObservationalSurveys}, \citealt{Gomez-Moran2011PostDistribution}). Progress towards an overarching and complete understanding of compact binary evolution across all areas discussed above relies on \textit{unbiased} samples of white dwarf binaries that are \textit{sufficiently large} to span the entire range of systems and evolutionary parameters (Fig.\,\ref{fig:schema}). Assembling these samples requires \textit{all-sky multi-epoch} spectroscopic surveys that will identify white dwarf binaries via the detection of emission lines and radial velocity variations. Low-resolution ($R\simeq5000$) optical spectroscopy is sensitive to short orbital period systems (hours to days), while systems with longer periods such as the progenitors of supersoft X-ray sources require high-resolution ($R\ge20\,000$) spectroscopy \citep[e.g.][]{Parsons2015TheBinary, Rebassa-Mansergas2017TheDR4}. 

\vspace{0.2cm}
\smallskip\noindent
\textbf{\large The need for ultraviolet spectroscopy}\\
All areas outlined above \textit{critically depend on access to high-resolution ($R\simeq20\,000-40\,000$) ultraviolet spectroscopy} to accurately characterize the white dwarfs in both interacting and detached binaries. Using the extremely precise \textit{Gaia} parallaxes, effective temperatures and masses of the white dwarfs can be measured from modelling the broad photospheric Lyman absorption lines. Thus, all other quantities that depend on these parameters can be obtained, e.g. mass accretion rates and angular momentum loss rates \citep{Townsley2009CataclysmicLoss}. Chemical abundances and rotation rates can be determined via spectral fitting with synthetic atmosphere models to the vast array of metal lines accessible in the far-ultraviolet, in particular transitions of carbon, silicon, and aluminum ($\simeq\,$900--2500\,\AA). High spectral resolution is required to resolve line blends and to distinguish photospheric and interstellar absorption. The study of disk wind structure will be possible by modelling P-Cygni profiles in the ultraviolet resonance lines and substructures tracing their variability. Post-supersoft X-ray sources will be identified via their anomalous N\,\textsc{v}/C\,\textsc{iv} abundance ratios. Detailed modelling of the photospheric abundances, which reflect the composition of the material accreted from the donor stars, will provide strong constraints on their current evolutionary states and on the initial configuration of these binaries. Very similar analyzes will allow us to distinguish the formation channels of individual AM\,CVn stars. Because these systems are extremely rare, and hence on average distant from Earth, such studies are beyond the reach of current instrumentation. The white dwarfs in the progenitors of supersoft sources are totally outshone by their massive companions at optical wavelengths, and ultraviolet spectroscopy is essential to measure their masses, and hence determine their future evolution. While \textit{FUSE} was limited in brightness and its resolution insufficient for the detailed analysis of the detected lines, \textit{IUE} and \textit{HST} spectroscopy was/is limited to $\lambda\ge1150$\,\AA, missing the bluer components of the Lyman series~--~extending the coverage down to the Lyman limit will significantly improve the robustness of the derived measurements, and is key to disentangle the white dwarf photospheric flux from contaminating emission originating in the accretion flow. Time-tagged photon detections are extremely powerful for the construction of arbitrary data cubes in wavelength and time to deal with the geocoronal airglow, orbital smearing, and intrinsic variability. Ideally, samples containing several hundred systems from each of the different populations described above will be observed, requiring the sensitivity to obtain high signal-to-noise ratio ($>$30) ultraviolet spectroscopy at a flux level of 1.2$\times$10$^{-16}$ erg/s/cm$^{2}$/$\AA$ at 1480\,\AA, typical for a white dwarf with a temperature of 12\,000\,K in systems as far as 500\,pc.


\vspace{0.2cm}
We conclude that future research on close binary evolution and thus the improvement of the current theoretical model relies on the combination of optical and ultraviolet spectroscopy. 

\clearpage

\bibliographystyle{mn_new}
\bibliography{references}
\end{document}